\documentclass[doublecol]{epl2}
\usepackage{amsmath,amssymb}

\begin{document}

\title{Firing regulation of fast-spiking interneurons by autaptic inhibition}

\author{Daqing Guo,\inst{1,2} Mingming Chen,\inst{1} Matja{\v z} Perc,\inst{3} Shengdun Wu,\inst{1} Chuan Xia,\inst{1} Yangsong Zhang,\inst{4} Peng Xu,\inst{1,2} Yang Xia,\inst{1,2} Dezhong Yao,\inst{1,2}}
\shortauthor{Guo et al.}
\institute{\inst{1}Key Laboratory for NeuroInformation of Ministry of Education, School of Life Science and Technology, University of Electronic Science and Technology of China, Chengdu 610054, People's Republic of China\\
\inst{2}Center for Information in Medicine, University of Electronic Science and Technology of China, Chengdu 610054, People's Republic of China\\
\inst{3}Faculty of Natural Sciences and Mathematics, University of Maribor, Koro{\v s}ka cesta 160, SI-2000 Maribor, Slovenia\\
\inst{4}School of Computer Science and Technology, Southwest University of Science and Technology, Mianyang, China}

\abstract{Fast-spiking (FS) interneurons in the brain are self-innervated by powerful inhibitory GABAergic autaptic connections. By computational modelling, we investigate how autaptic inhibition regulates the firing response of such interneurons. Our results indicate that autaptic inhibition both boosts the current threshold for action potential generation as well as modulates the input-output gain of FS interneurons. The autaptic transmission delay is identified as a key parameter that controls the firing patterns and determines multistability regions of FS interneurons. Furthermore, we observe that neuronal noise influences the firing regulation of FS interneurons by autaptic inhibition and extends their dynamic range for encoding inputs. Importantly, autaptic inhibition modulates noise-induced irregular firing of FS interneurons, such that coherent firing appears at an optimal autaptic inhibition level. Our result reveal the functional roles of autaptic inhibition in taming the firing dynamics of FS interneurons.}


\maketitle

In the mammalian brain, the majority of cortical inhibition is acquired by interneurons releasing gamma-aminobutyric acid (GABA) on principal cells~\cite{bartos2007synaptic}. This GABA-mediated inhibition is believed to play functional roles in many important neuronal computations, such as balancing excitation~\cite{brunel2000dynamics, jalan2015emergence, vogels2009gating, van1996chaos}, promoting neural oscillations~\cite{guo2012complex, dwivedi2014emergence, dwivedi2015optimization, singh2015synchronization, traub1997simulation, hajos2004spike} and gating multiple signals~\cite{Vogels1569}. As the most prominent type of interneurons, cortical fast-spiking (FS) interneurons have been observed to display highly variable electrophysiological properties~\cite{golomb2007mechanisms}. By integrating massive synaptic inputs, the FS interneurons process information and produce discrete trains of action potentials (so-called ``spike trains'') of their own. Exploring the firing regulation of FS interneurons is therefore fundamental for the understanding of complicated inhibition-related neuronal computations in the brain.

Moreover, a large proportion of FS interneurons in the neocortex and hippocampus have been identified to form self-feedbacks connections, termed as ``autapses'', onto themselves~\cite{Tamas15081997, Cobb1997629}. Previous electrophysiological recordings have revealed that most of FS interneuron autapses are mediated by GABA$_\text{A}$ receptors~\cite{connelly2010modulation} and, importantly, their transmission is discovered to be strong and robust~\cite{Bacci01022003, Bacci2006119}. These findings indicate that autaptic inhibition might have significant impacts on shaping the firing dynamics of FS interneurons. Such hypothesis has been confirmed by recent experimental and computational data, showing that the GABAergic autaptic inhibition might prevent repetitive firing, reduce firing rate and improve spike-timing precision~\cite{Bacci01022003, Bacci2006119, wang2014effect, wang2015}. However, so far it is still not completely established whether the autaptic inhibition participates into the modulations of input-output (IO) gain, firing pattern and firing regularity of FS interneurons.

In this Letter, we address this question by considering a FS interneuron driven by both external applied and autaptic inhibition currents. The dynamics of the FS interneuron is simulated using the Wang-Buzsaki (WB) neuron~\cite{Wang15101996}. As a paradigmatic model of the FS interneuron, the current balance equation of the WB neuron can be written as follows~\cite{Wang15101996}:
\begin{equation}
\begin{split}
C\frac{dV}{dt}=-I_{\text{Na}}-I_{\text{K}}-I_{\text{L}}+I_{\text{app}}+I_{\text{aut}},
\end{split}
\label{eq:1}
\end{equation}
where $V$ represents the membrane potential, and $I_{\text{Na}}=G_{\text{Na}}m_{\infty}^3h(V-E_{\text{Na}})$, $I_{\text{K}}=G_{\text{K}}n^4(V-E_{\text{K}})$ and $I_{\text{L}}=G_{\text{L}}(V-E_{\text{L}})$ are sodium, potassium and leakage currents through the membrane, respectively. Three gating variables obey the following equations:
\begin{equation}
\begin{split}
m_{\infty}&=\alpha_{m}/\left[\alpha_{m}+\beta_{m}\right],\\
\frac{dh}{dt}&=\phi\left[\alpha_{h}(1-h)-\beta_{h}h\right],\\
\frac{dn}{dt}&=\phi\left[\alpha_{n}(1-n)-\beta_{n}n\right].
\end{split}
\label{eq:2}
\end{equation}
Here $\alpha_x$ and $\beta_x$ ($x=m$, $h$ and $n$) are voltage-dependent rate functions given in~\cite{Wang15101996}. The applied external current is given by $I_{\text{app}}=I_{0}+\sigma\zeta(t)$, where $I_{0}$ is the mean current, $\zeta(t)$ is colored noise and $\sigma$ is the noise intensity. The colored noise is modeled as an Ornstein-Uhlenbeck process: $d\zeta(t)=-\frac{\zeta(t)}{\tau_c}dt +N_{\tau}\xi\sqrt{dt}$, where $\xi(t)$ is Gaussian white noise with zero mean and unit variance, $\tau_c=2.5$~ms is the correlation time and $N_{\tau}=(2/\tau_c)^{1/2}$ is a normalization constant such that $\zeta(t)$ has unit variance. We use the following parameters for the WB neuron~\cite{Wang15101996}: $C=1$~$\mu$F/cm$^2$, $G_{\text{Na}}=35$~mS/cm$^2$, $E_{\text{Na}}=55$~mV, $G_{\text{K}}=9$~mS/cm$^2$, $E_{\text{K}}=-90$~mV, $G_{\text{L}}=0.1$~mS/cm$^2$, $E_{\text{L}}=-65$~mV, and $\phi=5$. Under these conditions, the WB neuron has a small current threshold ($I_{\text{th}}=0.16$~$\mu$A/cm$^2$) in the absence of noise.

In our model, the autaptic inhibition current is conductance-based, given by:
\begin{equation}
\begin{split}
I_{\text{aut}}(t)= G_{\text{aut}}S(t)(E_{\text{syn}}-V),
 \end{split}
 \label{eq:3}
\end{equation}
where $G_{\text{aut}}$ is the autaptic coupling strength and $E_{\text{syn}}$ is the reversal potential. The synaptic variable $S$ is described by the classical first-order kinetic model~\cite{destexhe1994efficient, jeong2007synchrony, ermentrout2001effects}:
\begin{equation}
\begin{split}
\frac{dS}{dt}= \alpha[T](1-S)-\beta S,
 \end{split}
 \label{eq:4}
\end{equation}
where $\alpha$ and $\beta$ are forward and backward rate constants of GABA receptors, and $[T]$ represents the transmitter concentration. The relationship between the transmitter concentration and presynaptic voltage satisfies~\cite{destexhe1994efficient, jeong2007synchrony, ermentrout2001effects}:
\begin{equation}
\begin{split}
[T](V_{\text{pre}}) = \frac{T_{\text{max}}}{1+\exp[-(V_{\text{pre}}(t-\tau_{\text{d}}) - V_{p})/K_{p}]}.
 \end{split}
 \label{eq:5}
\end{equation}
Here $T_{\text{max}}$ is the maximal concentration of transmitter in the synaptic cleft, $V_{\text{pre}}$ denotes the presynaptic voltage, and $V_{\text{p}}$ and $K_{\text{p}}$ determine the threshold and the stiffness for the transmitter release, respectively. A parameter $\tau_{\text{d}}$ is introduced to mimic the autaptic transmission delay due to the finite propagation speed of action potentials. Unless otherwise stated, we set these synaptic-related parameters as~\cite{destexhe1994efficient, jeong2007synchrony, ermentrout2001effects}: $\alpha=2.0$~nM$^{-1}$ms$^{-1}$, $\beta=0.5$~nM$^{-1}$ms$^{-1}$ $E_{\text{syn}}=-80$~mV, $T_{\text{max}}=1$~nM, $V_{\text{p}}=-10$~mV, $K_{\text{p}}=10$~mV, and $\tau=1$~ms. This choice corresponds to an autapse mediated by GABA$_\text{A}$ receptors.

The above system is integrated by using the Euler-Maruyama algorithm with a fixed time step $h=0.01$~ms~\cite{kloeden2012numerical}. We use uniformly distributed values between -60 and 0~mV for the initial membrane potential to identify possible multistability. For each set of initial conditions we run simulations for 5000~ms to collect a sufficiently high number of spiking events for further analysis. For different measures used in this work, the final results are averaged over $50$ independent realizations.

\begin{figure}[!t]
\includegraphics[width=8.5cm]{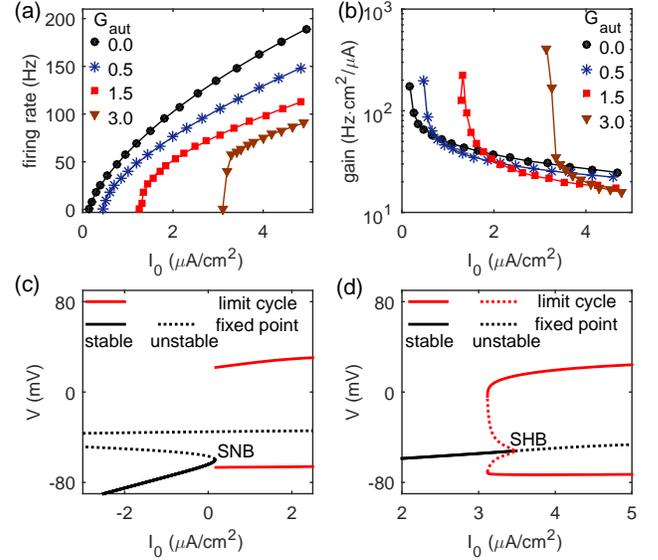}
\caption{(Color online)
Autaptic coupling strength modulates the firing response of FS interneurons in the absence of noise and transmission delay. (a) Dependence of output firing rate on $I_0$ at different self-feedback levels. (b) Input-output (IO) gain versus $I_0$ at different self-feedback levels. Autaptic coupling strengths in (a) and (b) are: $G_{\text{aut}}=0.0$, 0.5, 1.5 and 3.0~mS/cm$^2$. Bifurcation diagrams of the membrane potential $V$ reveal that the WB neuron undergoes a saddle-node bifurcation (SNB) at $I_{0}=0.16$~$\mu$A/cm$^2$ for $G_{\text{aut}}=0.0$~mS/cm$^2$ (c) and a subcritical Hopf bifurcation (SHB) at $I_{0}=3.46$~$\mu$A/cm$^2$ for $G_{\text{aut}}=3.0$~mS/cm$^2$ (d). Other parameters values are $\sigma=0$~$\mu$A/cm$^2$ and $\tau_{\text{d}}=0$~ms.}
\label{fig:1}
\end{figure}

We start by examining how autaptic inhibition impacts the firing response of FS interneurons in the absence of noise and transmission delay. Figure~\ref{fig:1}(a) shows the output firing rate of the WB model neuron versus the mean current $I_0$ at different autaptic coupling strengths. Due to negative self-feedback, increasing $G_{\text{aut}}$ shifts the current threshold of the output firing rate towards to the high current regime. Further quantitative analysis reveals that such current threshold is almost linear increase with increasing $G_{\text{aut}}$ (data not shown). Consistent with preceding research~\cite{wang2014effect, wang2015}, these results reveal a subtractive effect of autaptic inhibition on the output firing rate of FS interneurons.

To estimate how autaptic inhibition mediates the input-output relationship of FS interneurons, we calculate the IO gain, defined as the slope of firing rate curve at a specific mean current, versus the mean current at different self-feedback levels [Fig.~\ref{fig:1}(b)]. When autaptic inhibition is weak, the IO gain has a small value near and above the current threshold. In this case, the FS interneuron has a continuous firing rate curve and exhibits class I excitability [$G_{\text{aut}}=0.5$~mS/cm$^2$ in Fig.~\ref{fig:1}(a)]. As $G_{\text{aut}}$ increases, the IO gain at the initial stage of the firing rate curve becomes lager. For a sufficiently strong $G_{\text{aut}}$, the IO gain at the initial rising stage is so large that the FS interneuron has a discontinuous firing rate curve, indicating to display the class II excitability [$G_{\text{aut}}=3.0$~mS/cm$^2$ in Fig.~\ref{fig:1}(a)]. The bifurcation analysis confirms that strong self-feedback inhibition changes the intrinsic firing properties of the WB neuron from a saddle-node bifurcation (SNB) to a subcritical Hopf bifurcation (SHB) [Figs.~\ref{fig:1}(c) and \ref{fig:1}(d)]. As discussed in~\cite{Prescott2008}, these two bifurcations correspond to dynamical mechanisms for spike initiation in class I and II neurons, respectively. In experiments, the FS interneurons have been found to display either class I or class II excitability~\cite{Tateno2283,Ho18072012}, and our results might thus provide a biophysical basis for the understanding of this switch phenomenon. Moreover, we find that the IO gain is reduced as the mean current $I_0$ is increased [Fig.~\ref{fig:1}(b)]. Such reduction feature is physiologically important, because it restricts the explosive growth of the firing rate of FS interneurons. This reduction trend is enhanced with the increasing of $G_{\text{aut}}$ [see Fig.~\ref{fig:1}(b)]. As a result, the FS interneuron driven by a weaker autaptic inhibition has a relatively larger IO gain in the strong $I_0$ region. Overall, these above observations suggest that autaptic inhibition has a converging effect on firing rate in the weak mean current region and a diverging effect on firing rate in the strong mean current region. This might be because the closed-loop autaptic inhibition current shapes the firing susceptibility of FS interneurons in a highly nonlinear manner.

\begin{figure}[!t]
\centering
\includegraphics[width=8.5cm]{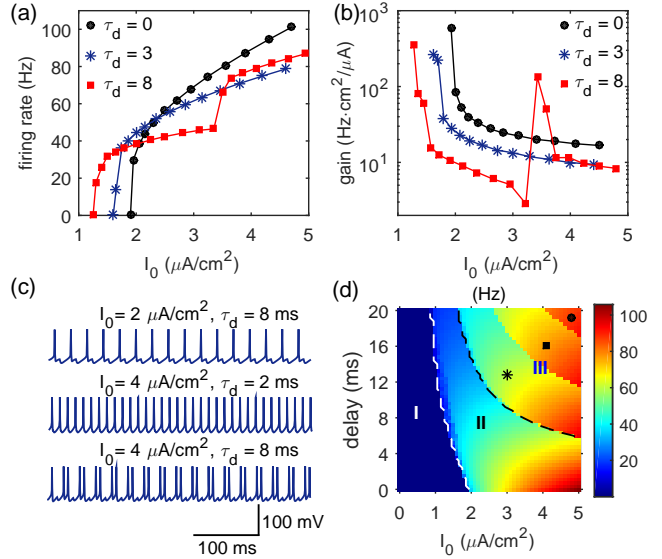}
\caption{(Color online) Autaptic transmission delay modulates both firing rate and pattern of FS interneurons in the absence of noise. (a) Dependence of output firing rate on mean current $I_0$ for different autaptic transmission delays. (b) IO gain is plotted versus mean current for different autaptic transmission delays. Autaptic transmission delays considered in (a) and (b) are: $\tau_{\text{d}}=0$, 3 and 8~ms. (c) Typical membrane potentials under different conditions. (d) Firing rate and pattern analysis in the $(I_0,\tau_{\text{d}})$ panel. Three firing patterns are observed: silent state (I), tonic firing (II), and burst-like firing (III). In (d), the dashed white line denotes the current threshold $I_{\text{th}}$ for different $\tau_{\text{d}}$, and the dashed black line represents the transition boundary between tonic firing and burst-like firing patterns. Symbols in region III denote the number of spikes contained in each periodic cycle: asterisk (2 spikes), square (3 spikes) and circle (4 spikes). Here we have used $\sigma =0$~$\mu$A/cm$^2$ and $G_{\text{aut}}=2.0$~mS/cm$^2$.}
\label{fig:2}
\end{figure}

We then turn to the autaptic transmission delay and investigate how delay parameter $\tau_{\text{d}}$ shapes the firing response of FS interneurons. In literature, there is a broad consensus that the synaptic transmission delay is an important intrinsic property of neural processing and cannot be arbitrarily ignored in modelling studies \cite{tass1996delay, popovych2005effective, masoller2008interplay, wang2009synchronization, Wessel2009Detection, li2009synchronization, perlikowski2010periodic, tang2011delay, singh2013role}. Figure~\ref{fig:2}(a) depicts the output firing rate of FS interneurons as a function of the mean current $I_0$ for different autaptic transmission delays. We find that increasing the delay parameter $\tau_{\text{d}}$ postpones the effect of autaptic inhibition current, thus shifting the current threshold towards to the low current region [Fig.~\ref{fig:2}(a), and also see the white dashed line in Fig.~\ref{fig:2}(d)]. This indicates that FS interneurons with a longer self-feedback delay tend to have a relatively smaller current threshold for spiking.

Indeed, our results presented in Fig.~\ref{fig:2}(a) also reveal that the autaptic transmission delay strongly influences the shape of firing rate curve. As we can see, increasing the autaptic transmission delay makes the firing rate curve to become more flat after an initial stage of steep growth. This leads to a relatively small IO gain for a long autaptic transmission delay at least in the intermediate mean current region [Fig.~\ref{fig:2}(b)]. For a sufficiently long $\tau_{\text{d}}$, we observe that a sudden ``jump'' of firing rate occurring in the strong mean current region. Consistently, the similar sudden jump can be also discovered in the corresponding IO gain curve [$\tau_{\text{d}}=8$~ms in Fig.~\ref{fig:2}(b)]. To understand the underlying mechanism of this sudden jump, several typical membrane potentials of the FS interneuron under different conditions are plotted in Fig.~\ref{fig:2}(c). By comparison, we find that such sudden jump is caused by the firing state transition of the FS interneuron from tonic firing pattern to burst-like firing pattern. For sufficiently long $\tau_{\text{d}}$ and strong $I_0$, the FS interneuron has enough time to fire more than once during a whole periodic cycle, before the inhibitory autaptic current caused by the first spike within the same periodic cycle starts to suppress its firing. Thus, the longer the autaptic transmission delay $\tau$, the larger the number of spikes that can be contained in each periodic cycle. This explanation is supported by performing two-dimensional firing rate and pattern analysis in the $(I_0,\tau_{\text{d}})$ panel [region III in Fig.~\ref{fig:2}(d)]. In agreement with the results reported in \cite{wang2014effect, wang2015}, the firing rate of the FS interneuron can be down-regulated if the value of $I_0$ is not too small, but such effects depend significantly on $\tau_{\text{d}}$. Together, these results emphasize that autaptic transmission delay not only modulates the firing response of FS interneurons, but also regulates their firing patterns.

\begin{figure}[!t]
\centering
\includegraphics[width=8.5cm]{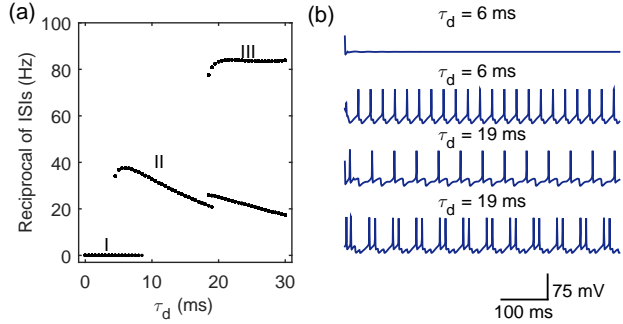}
\caption{(Color online) The FS interneuron with delayed autaptic inhibition exhibits multistable firing patterns. (a) Bifurcation diagram of the reciprocal of inter-spike intervals (ISIs) versus the autaptic transmission delay $\tau_{\text{d}}$. 
The FS interneuron may exhibit three firing patterns: silent state (I), tonic firing (II), and burst-like firing (III).
(b) Typical membrane potentials under different conditions. From top to bottom, the considered autaptic transmission delays are $\tau_{\text{d}}=6$, 6, 19 and 19~ms, and the chosen initial membrane potentials are $V=-13.73$, -33.53, -5.33 and -57.81~mV. Other parameter values are $I_0=2.0$~$\mu$A/cm$^2$, $\sigma =0$~$\mu$A/cm$^2$, $G_{\text{aut}}=3.0$~mS/cm$^2$ and $\tau_{\text{d}}=1$~ms.}
\label{fig:3}
\end{figure}

A more detailed bifurcation analysis reveals that multistable firing patterns can also arise in the considered dynamical system~\cite{Foss1996, Foss2000}. As shown in Fig.~\ref{fig:3}(a), the FS interneuron is either in the silent state or exhibits tonic firing for shorter $\tau_{\text{d}}$ between 4.5 and 8.5 ms [$\tau_{\text{d}}=6$~ms in Fig.~\ref{fig:3}(b)], whereas it displays either tonic firing or burst-like firing for longer $\tau_{\text{d}}$ between 18 and 19.5 ms [$\tau_{\text{d}}=19$~ms in Fig.~\ref{fig:3}(b)]. The first multistability, occurring in the small autaptic transmission delay region, is associated with the bistability that is due to a subcritical Hopf bifurcation, while the second multistability is a consequence of the co-existence of multiple limit cycle attractors that emerge at appropriately long autaptic transmission delays~\cite{Foss1996, Foss2000}. When FS interneurons operate in these multistable regions the final firing pattern thus depends on the initial conditions, and moreover, fluctuations may induce a switch between different firing patterns.

\begin{figure}[!t]
\centering
\includegraphics[width=8.5cm]{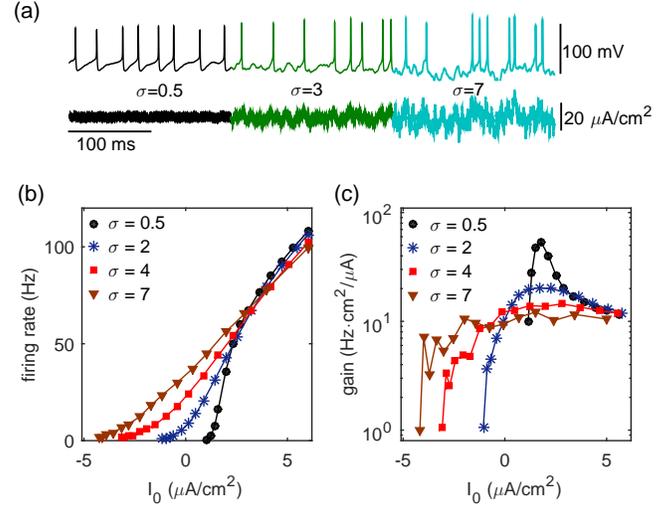}
\caption{(Color online) Neuronal noise impacts the firing regulation of FS interneurons by autaptic inhibition. (a) Typical membrane potentials (top) and corresponding applied currents (bottom) at different noise levels. Here we set $I_{0}=2.0$~$\mu$A/cm$^2$, and choose noise intensities are $\sigma =0.5$, 3 and 7~$\mu$A/cm$^2$. (b) Dependence of output firing rate on mean current $I_0$ at different noise levels. (c) IO gain is plotted versus mean current at different noise levels. In (b) and (c), the considered noise intensities are $\sigma =0.5$, 2, 4 and 7~$\mu$A/cm$^2$. Here we have used $G_{\text{aut}}=2.0$~mS/cm$^2$ and $\tau_{\text{d}}=1$~ms.}
\label{fig:4}
\end{figure}

Because real neurons work in noisy environments~\cite{neiman1994stochastic, pikovsky1997coherence, anishchenko1999stochastic, ma2010collective, Ma2011, destexhe2012neuronal, qin2014Autapse, PhysRevE91052920, guo2009stochastic}, we next introduce a certain level of neuronal noise to our system. This gives rise to stochastic fluctuations in the membrane potential, and at sufficiently strong intensities, neuronal noise induces irregular spiking [Fig.~\ref{fig:4}(a)]. To investigate how neuronal noise alters the firing regulation of FS interneurons by autaptic inhibition, we plot the output firing rate of FS interneurons and corresponding IO gain versus the mean current $I_0$ at different noise levels in Figs.~\ref{fig:4}(b) and \ref{fig:4}(c), respectively. As we can see, increasing the noise intensity shifts the current threshold towards to the low current region. For a sufficiently high noise level, the FS interneuron can emit spikes even when $I_0$ is negative [Fig.~\ref{fig:4}(b)]. This makes the firing rate curves flatter at strong noise intensity, thus yielding a relatively smaller IO gain at the initial rising stage [for example, see $\sigma =7$ and 7~$\mu$A/cm$^2$ in Fig.~\ref{fig:4}(c)]. As $I_0$ is increased, the firing rate curves for different noise levels gradually approach each other. By analyzing the IO gain, we identify that with an increasing mean current the firing rate curves grow slowly at first, then the growth accelerates, and finally there is slowing down again, especially for weak noise intensities [Fig.~\ref{fig:4}(c)]. In the strong mean current region, the firing of FS interneurons becomes so insensitive to neuronal noise that there are no significant qualitative differences between the firing rate curves [Fig.~\ref{fig:4}(b)]. These noise-induced effects modulate the intrinsic integration property of FS interneurons, conferring them a relatively broad dynamic range for encoding inputs under strong noise condition [see $\sigma =7$~$\mu$A/ms$^2$ in Fig.~\ref{fig:4}(b)]. We note that in~\cite{Khubieh24072015} experimental and computational research revealed a similar noise-induced broad dynamic range of inputs for hippocampal pyramidal neurons, and it was argued that the rate of coding might benefit from such a broad dynamic range. In view of our results, stochastic fluctuations might thus also facilitate the information integration capability of FS interneurons in the brain.

\begin{figure}[!t]
\centering
\includegraphics[width=8.5cm]{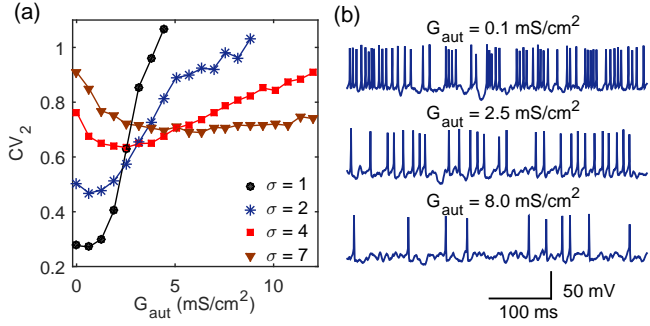}
\caption{(Color online) Autaptic inhibition modulates the irregular firing of FS interneurons. (a) CV$_{\text{2}}$ value is plotted as a function of autaptic coupling strength $G_{\text{aut}}$ for different noise levels. The considered noise intensities are $\sigma =1$, 2, 4 and 7~$\mu$A/cm$^2$. (b) Typical membrane potentials of the FS interneuron at different autaptic coupling strengths, with a fixed noise intensity $\sigma =4$~$\mu$/cm$^2$. From top to bottom, the autaptic coupling strengths are $G_{\text{aut}}=0.1$, 2.5 and 8.0~mS/cm$^2$. Here we have used $I_{0}=2.0$~$\mu$A/cm$^2$ and $\tau_{\text{d}}=1$~ms.}
\label{fig:5}
\end{figure}

Finally, we ask whether autaptic inhibition modulates the irregular firing generated by FS interneurons. To address this, we employ the local coefficient of variation of ISIs ($\text{CV}_2$) to quantify the temporal regularity of a spike train~\cite{Holt1996}. Mathematically, the $\text{CV}_2$ is defined as~\cite{Holt1996} $\text{CV}_2=2\langle |T_{i}-T_{i+1}| \rangle/\langle T_{i}+T_{i+1} \rangle$, where the symbol $\langle\cdot\rangle$ represents the average over time, $T_i = t_{i+1}-t_i$, and $t_i$ is the time of the $i$-th neuronal firing. By definition, a smaller CV$_{2}$ value corresponds to a better spiking regularity. Figure~\ref{fig:5}(a) depicts the CV$_{2}$ value as a function of the autaptic coupling strength $G_{\text{aut}}$ for different noise levels. As $G_{\text{aut}}$ increases, we find that each CV$_{2}$ curve first drops and then rises, and the smallest CV$_{2}$ value is achieved at an optimal $G_{\text{aut}}$. When autaptic inhibition is weak, the FS interneuron emits spikes in a high rate and neuronal noise induced stochastic fluctuations drive it to fire bursts occasionally [top panel in Fig.~\ref{fig:5}(b)]. Under this condition, the spike train generated by the FS interneuron has a low temporal regularity. For an appropriate self-feedback level, the autaptic inhibition acts as an effective low-pass filter~\cite{wang2015}, which not only suppresses burst firing, but also ensures that the FS interneuron fires well-separated spikes with a relatively high temporal regularity [middle panel in Fig.~\ref{fig:5}(b)]. However, too strong autaptic inhibition greatly depresses neuronal activity, resulting that the FS interneuron only produces few scattered spikes with a low temporal regularity [bottom panel in Fig.~\ref{fig:5}(b)]. We also note that the delayed feedback due to diffusion coupling has been found to increase coherent motion in various excitable systems~\cite{Prager2007, Goldobin2003, Janson2004}. Our findings further contribute to these results by demonstrate that autaptic inhibition mediated by chemical receptors can modulate the firing regularity of FS interneurons, and that suitable tuning of the autaptic coupling strength may trigger coherent firing.

In conclusion, we have systematically investigated how autaptic inhibition regulates the firing response of FS interneurons. By computational modelling, we identified that autaptic inhibition not only enhances the current threshold for action potential generation, but also contributes to the gain control of FS interneurons. In particular, we showed that autaptic inhibition has a converging effect on firing rate in the weak mean current region, and divergently modulate firing rate in the strong mean current region. Strong autaptic inhibition can switch the intrinsic excitability of FS interneurons by changing their bifurcation structure. Further investigation demonstrated that autaptic transmission delay also participates in the firing regulation of FS interneurons. Increasing the autaptic transmission delay was found to cause dynamical transition for FS interneurons from tonic firing pattern to burst-like firing pattern. By suitably tuning the autaptic transmission delay, the FS interneurons may produce multistable firing patterns. Moreover, we observed that both the firing rate and the input-output gain are shaped by neuronal noise.Under strong noise condition, the FS interneuron has a broad dynamic range for encoding inputs, thus exhibiting a strong information integration capability. Remarkably, we uncovered that autaptic inhibition modulates neuronal noise-induced irregular firing of FS interneurons, and the coherent firing occurs at an optimal autaptic coupling strength.

In terms of biological implications, we firstly note that neurons process information through integration of synaptic inputs from dendrites~\cite{Greg2015, Songting2014}. Experimental studies have revealed that dendritic integration is highly nonlinear~\cite{Greg2015, Songting2014}. The temporal link between a spike and a following large GABAergic conductance makes autaptic inhibition an important candidate in modulating the dendritic integration of FS interneurons. Theoretically, self-innervation by an inhibitory autapse might shape the nonlinear dendritic integration, through controlling the timing of dendritic spikes, preventing excessive excitation, and taming firing patterns of dendritic spikes. Secondly, autpatic inhibition may play dual roles in the generation of burst firing. The short delayed autaptic inhibition reduces the membrane potential of the FS interneuron that just fired, thus providing a biophysical basis to suppress burst firing. In contrast, autaptic inhibition with a larger transmission delay may offer a sufficiently long time-window to the FS interneuron driven by strong external currents, thus enabling it to produce burst-like firings.

Our results emphasize the functional significance of autaptic inhibition in mediating neurodynamics, and provide computational evidence to rule out an old developmental accident hypothesis concerning autapses~\cite{Bekkers1998R52,Lubke15051996}. We hope that our research will inspire testable hypotheses for electrophysiological experiments in the near future. We conclude by noting that autapses have also been observed in pyramidal neurons
\cite{Lubke15051996}, and that thus the role of autaptic excitation in firing regulation in these neurons also merits further research.

We thank Prof.~Ying Liu and Dr.~Hongzhi You for the careful proofreading of our manuscript. This research is supported by the National Natural Science Foundation of China (Grant Nos.~81571770, 61527815, 61201278, 81371636, 81401484 and 81330032) and the Slovenian Research Agency (Grants P5-0027 and J1-7009).


\begin{thebibliography}{10}
\expandafter\ifx\csname url\endcsname\relax\def\url#1{\texttt{#1}}\fi

\bibitem{bartos2007synaptic}
\Name{Bartos M., Vida I. \and Jonas P.} \REVIEW{Nat Rev Neurosci}{8}{2007}{45}.

\bibitem{brunel2000dynamics}
\Name{Brunel N.} \REVIEW{J. Comput. Neurosci.}{8}{2000}{183}.

\bibitem{jalan2015emergence}
\Name{Jalan S. \and Dwivedi S.~K.} \REVIEW{EPL}{112}{2015}{48003}.

\bibitem{vogels2009gating}
\Name{Vogels T.~P. \and Abbott L.} \REVIEW{Nat. Neurosci.}{12}{2009}{483}.

\bibitem{van1996chaos}
\Name{van Vreeswijk C. \and Sompolinsky H.} \REVIEW{Science}{274}{1996}{1724}.

\bibitem{guo2012complex}
\Name{Guo D., Wang Q. \and Perc M.} \REVIEW{Phys. Rev. E}{85}{2012}{061905}.

\bibitem{dwivedi2014emergence}
\Name{Dwivedi S.~K. \and Jalan S.} \REVIEW{Phys. Rev. E}{90}{2014}{032803}.

\bibitem{dwivedi2015optimization}
\Name{Dwivedi S.~K., Sarkar C. \and Jalan S.} \REVIEW{EPL}{111}{2015}{10005}.

\bibitem{singh2015synchronization}
\Name{Singh A., Ghosh S., Jalan S. \and Kurths J.} \REVIEW{EPL}{111}{2015}{30010}.

\bibitem{traub1997simulation}
\Name{Traub R.~D., Jefferys J.~G. \and Whittington M.~A.} \REVIEW{J. Comput. Neurosci.}{4}{1997}{141}.

\bibitem{hajos2004spike}
\Name{H{\'a}jos N., P{\'a}lhalmi J., Mann E.~O., N{\'e}meth B., Paulsen O. \and
  Freund T.~F.} \REVIEW{J. Neurosci.}{24}{2004}{9127}.

\bibitem{Vogels1569}
\Name{Vogels T.~P., Sprekeler H., Zenke F., Clopath C. \and Gerstner W.}
  \REVIEW{Science}{334}{2011}{1569}.

\bibitem{golomb2007mechanisms}
\Name{Golomb D., Donner K., Shacham L., Shlosberg D., Amitai Y. \and Hansel D.}
  \REVIEW{PLoS Comput. Biol.}{3}{2007}{e156}.

\bibitem{Tamas15081997}
\Name{Tam{\'a}s G., Buhl E.~H. \and Somogyi P.} \REVIEW{J. Neurosci.}{17}{1997}{6352}.

\bibitem{Cobb1997629}
\Name{Cobb S., Halasy K., Vida I., Ny{\'i}ri G., Tam{\'a}s G., Buhl E. \and
  Somogyi P.} \REVIEW{Neurosci.}{79}{1997}{629 }.

\bibitem{connelly2010modulation}
\Name{Connelly W.~M. \and Lees G.} \REVIEW{J. Physiol.}{588}{2010}{2047}.

\bibitem{Bacci01022003}
\Name{Bacci A., Huguenard J.~R. \and Prince D.~A.} \REVIEW{J Neurosci.}{23}{2003}{859}.

\bibitem{Bacci2006119}
\Name{Bacci A. \and Huguenard J.~R.} \REVIEW{Neuron}{49}{2006}{119}.

\bibitem{wang2014effect}
\Name{Wang H.~T., Wang L.~F., Chen Y.~L. \and Chen Y.} \REVIEW{Chaos}{24}{2014}(033122).


\bibitem{wang2015}
\Name{Wang H.~T. \and Chen Y.} \REVIEW{Chin. Phys. B}{24}{2015}(128709).


\bibitem{Wang15101996}
\Name{Wang X.-J. \and Buzs¨¢ki G.} \REVIEW{J. Neurosci.}{16}{1996}{6402}.

\bibitem{destexhe1994efficient}
\Name{Destexhe A., Mainen Z.~F. \and Sejnowski T.~J.} \REVIEW{Neural Computation}{6}{1994}{14}.

\bibitem{jeong2007synchrony}
\Name{Jeong H.~Y. \and Gutkin B.} \REVIEW{Neural Computation}{19}{2007}{706}.

\bibitem{ermentrout2001effects}
\Name{Ermentrout B., Pascal M. \and Gutkin B.} \REVIEW{Neural Computation}{13}{2001}{1285}.

\bibitem{kloeden2012numerical}
\Name{Kloeden P.~E., Platen E. \and Schurz H.} \Book{Numerical solution of SDE
  through computer experiments} (Springer-Verlag, Berlin) 1994.

\bibitem{Prescott2008}
\Name{Prescott S.~A., De Koninck Y. \and Sejnowski T.~J.} \REVIEW{PLoS Comput. Biol.}{4}{2008}{e1000198}.

\bibitem{Tateno2283}
\Name{Tateno T., Harsch A. \and Robinson H. P.~C.} \REVIEW{J. Neurophysiol.}{92}{2004}{2283}.

\bibitem{Ho18072012}
\Name{Ho E. C.~Y., Str¨¹ber M., Bartos M., Zhang L. \and Skinner F.~K.}
  \REVIEW{J. Neurosci.}{32}{2012}{9931}.

\bibitem{tass1996delay}
\Name{Tass P., Kurths J., Rosenblum M.~G., Guasti G. \and Hefter H.}
  \REVIEW{Phys. Rev. E}{54}{1996}{R2224}.

\bibitem{popovych2005effective}
\Name{Popovych O.~V., Hauptmann C. \and Tass P.~A.} \REVIEW{Phys. Rev. Lett.}{94}{2005}{164102}.

\bibitem{masoller2008interplay}
\Name{Masoller C., Torrent M.~C. \and Garc\'{\i}a-Ojalvo J.} \REVIEW{Phys. Rev. E}{78}{2008}{041907}.

\bibitem{wang2009synchronization}
\Name{Wang Q., Perc M., Duan Z. \and Chen G.} \REVIEW{Phys. Rev. E}{80}{2009}{026206}.

\bibitem{Wessel2009Detection}
\Name{Wessel N., Suhrbier A., Riedl M., Marwan N., Malberg H., Bretthauer G.,
  Penzel T. \and Kurths J.} \REVIEW{EPL}{87}{2009}{10004}.

\bibitem{li2009synchronization}
\Name{LI Y.-L., MA J., LIU Y.-J., ZHANG L.-P. \and SU J.-Y.}
  \REVIEW{Int. J. Mod. Phys. C}{20}{2009}{1521}.

\bibitem{perlikowski2010periodic}
\Name{Perlikowski P., Yanchuk S., Popovych O.~V. \and Tass P.~A.} \REVIEW{Phys. Rev. E}{82}{2010}{036208}.

\bibitem{tang2011delay}
\Name{Tang J., Ma J., Yi M., Xia H. \and Yang X.} \REVIEW{Phys. Rev. E}{83}{2011}{046207}.

\bibitem{singh2013role}
\Name{Singh A., Jalan S. \and Kurths J.} \REVIEW{Phys. Rev. E}{87}{2013}{030902}.

\bibitem{Foss1996}
\Name{Foss J., Longtin A., Mensour B. \and Milton J.} \REVIEW{Phys. Rev. Lett.}{76}{1996}{708}.

\bibitem{Foss2000}
\Name{Foss J. \and Milton J.} \REVIEW{J. Neurophysiol.}{84}{2000}{985}.

\bibitem{neiman1994stochastic}
\Name{Neiman A. \and Schimansky-Geier L.} \REVIEW{Phys. Rev. Lett.}{72}{1994}{2988}.

\bibitem{pikovsky1997coherence}
\Name{Pikovsky A.~S. \and Kurths J.} \REVIEW{Phys. Rev. Lett.}{78}{1997}{775}.

\bibitem{anishchenko1999stochastic}
\Name{Anishchenko V.~S., Neiman A.~B., Moss F. \and Shimansky-Geier L.} \REVIEW{Physics-Uspekhi}{42}{1999}{7}.

\bibitem{ma2010collective}
\Name{MA J., ZHANG A.-H., TANG J. \and JIN W.-Y.} \REVIEW{J. Biol. Syst.}{18}{2010}{243}.

\bibitem{Ma2011}
\Name{Ma J., Wu Y., Ying H. \and Jia Y.} \REVIEW{Chinese Sci. Bull.}{56}{2011}{151}.

\bibitem{destexhe2012neuronal}
\Name{Destexhe A. \and Rudolph-Lilith M.} \Book{Neuronal noise} (Springer Science \& Business Media) 2012.

\bibitem{qin2014Autapse}
\Name{Qin H., Ma J., Wang C. \and Wu Y.} \REVIEW{PLoS ONE}{9}{2014}{e100849}.

\bibitem{PhysRevE91052920}
\Name{Bashkirtseva I., Neiman A.~B. \and Ryashko L.} \REVIEW{Phys. Rev. E}{91}{2015}{052920}.

\bibitem{guo2009stochastic}
\Name{Guo D. \and Li C.} \REVIEW{Phys. Rev. E }{79}{2009}{051921}.

\bibitem{Khubieh24072015}
\Name{Khubieh A., Ratt{\'e} S., Lankarany M. \and Prescott S.~A.} \REVIEW{Cereb. Cortex}{2015}{in print}.

\bibitem{Holt1996}
\Name{Holt G.~R., Softky W.~R., Koch C.~D. \and Douglas R.~J.} \REVIEW{J Neurophysiol.}{75}{1996}{1806}.

\bibitem{Prager2007}
\Name{Prager T., Lerch H.~P., Schimansky-Geier L. \and Scholl E.} \REVIEW{J. Phys. A}{40}{2007}{11045}.

\bibitem{Goldobin2003}
\Name{Goldobin D., Rosenblum M. \and Pikovsky A.} \REVIEW{Phys. Rev. E}{67}{2003}{061119}.

\bibitem{Janson2004}
\Name{Janson N.~B., Balanov A.~G. \and Scholl E.} \REVIEW{Phys. Rev. Lett.}{93}{2004}{010601}.

\bibitem{Greg2015}
\Name{Stuart G.~ J. \and Spruston N.} \REVIEW{Nat. Neurosci.}{18}{2015}{1713}.

\bibitem{Songting2014}
\Name{Li S.~T., Liu N., Zhang X.~H, Zhou D. \and Cai D.} \REVIEW{PLoS Comput. Biol.}{10}{2014}{e1004014}.

%

\bibitem{Bekkers1998R52}
\Name{Bekkers J.~M.} \REVIEW{Curr. Biol.}{8}{1998}{R52}.

\bibitem{Lubke15051996}
\Name{L{\"u}bke J., Markram H., Frotscher M. \and Sakmann B.} \REVIEW{J. Neurosci.}{16}{1996}{3209}.

\end{thebibliography}

\end{document}